Land Use and Land Cover Classification

Using Deep Learning Techniques

by

Nagesh Kumar Uba

A Thesis Presented in Partial Fulfillment
of the Requirements for the Degree
Master of Science

Approved April 2016 by the
Graduate Supervisory Committee:

John Femiani, Chair
Anshuman Razdan
Ashish Amresh

ARIZONA STATE UNIVERSITY

May 2016

ABSTRACT

Large datasets of sub-meter aerial imagery represented as orthophoto mosaics are widely available today, and these data sets may hold a great deal of untapped information. This imagery has a potential to locate several types of features; for example, forests, parking lots, airports, residential areas, or freeways in the imagery. However, the appearances of these things vary based on many things including the time that the image is captured, the sensor settings, processing done to rectify the image, and the geographical and cultural context of the region captured by the image. This thesis explores the use of deep convolutional neural networks to classify land use from very high spatial resolution (VHR), orthorectified, visible band multispectral imagery. Recent technological and commercial applications have driven the collection a massive amount of VHR images in the visible red, green, blue (RGB) spectral bands, this work explores the potential for deep learning algorithms to exploit this imagery for automatic land use/ land cover (LULC) classification. The benefits of automatic visible band VHR LULC classifications may include applications such as automatic change detection or mapping. Recent work has shown the potential of Deep Learning approaches for land use classification; however, this thesis improves on the state-of-the-art by applying additional dataset augmenting approaches that are well suited for geospatial data. Furthermore, the generalizability of the classifiers is tested by extensively evaluating the classifiers on unseen datasets and we present the accuracy levels of the classifier in order to show that the results actually generalize beyond the small benchmarks used in training. Deep networks have many parameters, and therefore they are often built with very large sets of labeled data. Suitably large datasets for LULC are not easy to come by, but techniques such as refinement learning allow networks trained for one task to be retrained to perform another recognition task. Contributions of this thesis include demonstrating that deep networks trained for image recognition in one task (ImageNet) can be efficiently transferred to remote sensing applications and perform as well or better than manually crafted classifiers without requiring massive training data sets. This is demonstrated on the UC Merced dataset, where 96% mean accuracy is achieved using a CNN (Convolutional Neural Network) and 5-fold cross validation. These results are further tested on unrelated VHR images at the same resolution as the training set.

# DEDICATION

*Dedicated to my mom*

ACKNOWLEDGMENTS

I am using this opportunity to express my sincere gratitude to my mentors Dr. John Femiani and Dr. Anshuman Razdan, who supported me throughout the course of the Masters thesis and without them, none of this could have happened. I am thankful for their aspiring guidance, friendly advice and invaluably constructive criticism during the status review meetings and the project work. I am sincerely grateful to them for sharing their truthful and illuminating views on some issues related to the research. I express my warm thanks for their constant support and guidance in this research.

I would also like to thank Mr. Michael Katic a Senior Software Engineer at the I3DEA labs, ASU Polytechnic and all the people who provided me with the facilities being required to conduct the experiments at the I3DEA lab.

Most importantly, none of this could have happened without my family. My mother who offered her encouragement through phone calls despite my limited devotion to correspondence. This dissertation stands as a testament to your unconditional love and support.

# TABLE OF CONTENTS

LIST OF TABLES

LIST OF FIGURES

PREFACE

As a beginning graduate student I had the opportunity to see Dr. John Femiani and Dr. Anshuman Razdan present the research work at the Image & 3D Data Exploitation and Analysis (I3DEA) labs at the ASU Polytechnic campus. I was utterly fascinated by the kind of work they were doing. They helped me to explore an idea to categorize aerial images and extract features from these images. The approach started by trying to extract features from aerial images using traditional techniques, but these methods gave poor or fair results at best. Machine Learning and Deep Learning techniques showed promising results in the past Large Scale Visual Recognition Challenge (ILSVRC) ("ImageNet Large Scale Visual Recognition Competition (ILSVRC)," n.d.) competition. The Deep Learning library Caffe (Jia et al., 2014) is a high performing tool that I could quickly get my hands on and start trying the ideas. The approach is to design and come up with a high-performance Deep Learning classifier that does the job and at the same time quick and easy to build. Thus, the ideas of the approach of transfer learning were suggested. I am thankful that by the guidance of my mentors Dr. Femiani and Dr. Razdan, I have achieved this feat by doing Land Use Land Cover classification with UC Merced ("UC Merced Land Use Dataset," n.d.) (O. A. Penatti, Nogueira, & dos Santos, 2015) dataset and tested the classifier with unrelated random samples.

CHAPTER 1

INTRODUCTION

Recent technological advancements in remote sensing and high-resolution image capturing by satellite drones and airplanes have been a huge help in gathering datasets for research and development. High-resolution orthoimagery datasets are readily available for download by resources such as United States Geological Survey. With the abundance of the data, the question arises of how to make use of the data for technological applications such as civil engineering, environmental monitoring or data generation for simulation and training. There have been significant advances in remote sensing and high-resolution image processing, and a variety of Land Use and Land Cover(LULC) classification algorithms have been developed in the recent past. One concern about LULC classification is that at high resolutions, there is a significant amount of variability in the data. The benchmark datasets used to test and evaluate classification algorithms may not capture enough of the variability to generalize to unseen examples that may have been acquired at different times or locations.

Recent advances in machine learning have been accomplished through an approach called "deep learning". Deep learning refers to artificial neural networks comprised of many layers of artificial neurons, called perceptrons. Historically the number of layers in a neural network has been limited because the algorithms used to train networks became unstable when the depth (number of layers) of the network is increased. Recent advances in hardware as well as the availability of very large datasets and robust training algorithms have made deep neural networks not only tractable, but they outperform other algorithms by an order of magnitude in some problems ("ImageNet Large Scale Visual Recognition Competition (ILSVRC)," n.d.). One area where deep learning is particularly successful is computer vision, which uses a type of network called a "convolutional neural network", often called a ConvNet or CNN. Convolutional neural networks have the interesting property that the first layers of the network tend to learn patterns that mimic those observed in human vision (Gabor Wavelets, (Daugman, 1988)).

Deep learning frameworks usually need large sets of labeled data to train and classify the images. High-resolution orthoimagery is readily available, but for supervised-learning one first has to label the datasets to train the classifier. For example, the IMAGENET Large Scale Visual Recognition Challenge (ILSVRC) uses 10+ million images belonging to 400+ unique scene categories ("ImageNet Large Scale Visual Recognition Competition (ILSVRC)," n.d.). The challenge data is divided into 8.1M images for training, 20K images for validation and 381K images for testing coming from 401 scene categories.

For the ILSVRC-2012 object localization challenge, the validation and test data consisted of 150K photographs, collected from Flickr and other search engines, hand labeled with the presence or absence of 1000 object categories. The training data for object localization was the subset of ImageNet consisting of 1.2M images used as training data with 1000 categories. Generating such a large amount of annotated data is labor-intensive and takes a large amount of time. It is painstakingly time-consuming for humans to label manually each of the images from any new dataset.

Deep networks require large amount of data or they risk becoming overly specialized, however the process of generating labeled data for each new recognition task is expensive. One approach to solve this problem is to start from a pre-trained convolutional neural network, and use a technique called "refinement" or "transfer learning" to adapt the network to a new task. Transfer learning uses the existing pre-trained classifier and learns only on the top, fully connected layers, of the network.

Artificial neural networks are biologically inspired systems inspired by a human brain (Hopfield, 1988). The processing part of the human brain consists of billions of neurons and each neuron receives information from thousands of other neurons. A neuron can be studied as an input/output device. Neurons fundamentally transmit pulse-coded information. In an ANN (Artificial Neural Network) the input and output of this pulse coded system is a non-linear (sigmoid) function as shown in Figure 1. A sigmoid function is a non-linear function that is saturated at both ends. It is a bounded differentiable real function that is defined for all real input values and has a positive derivative at every point, and this is crucial to neural networks' computational properties (Mira & Sandoval, 1995).

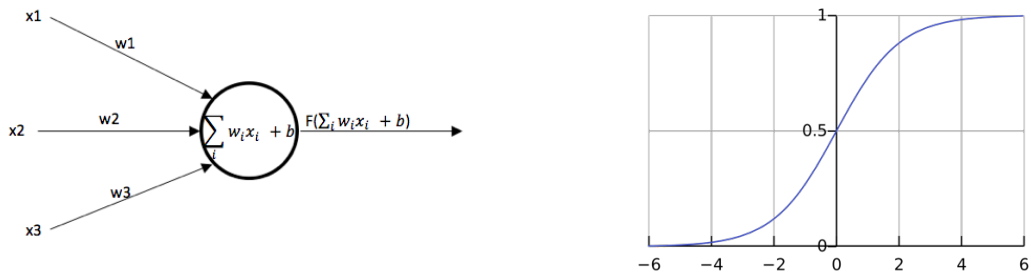

*Figure 1. Left:* A mathematical model of a biological neuron. *Right:* Sigmoid function

A perceptron is an artificial neural network element that is analogous to a biological neuron. An artificial neural network is an interconnected group of nodes that show a similar character to the vast network of neurons in a biological brain. In Figure 2, each node (circle) represents an artificial neuron, and an arrow represents a connection from the output of one neuron to the input of another (Strickland, 2015).

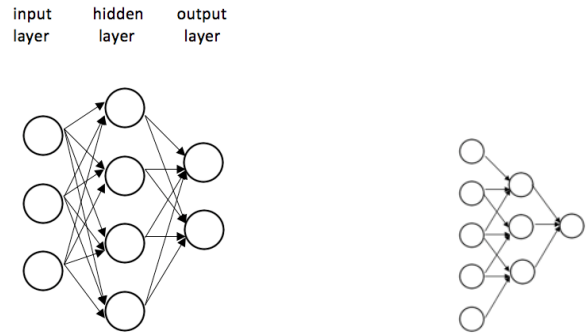

*Figure 2. Left:* A Multi layer perceptron with a single hidden layer *Right:* Model of a Convolutional Neural Network

A typical neural network consists of one or two hidden layers of neurons feeding one another. Deep learning neural networks have many hidden layers. Before 2006, attempts at training deep architectures have not shown much success. A revolutionary work published in 2006-2007 by (Hinton, Osindero, & Teh, 2006), (Bengio, Lamblin, Popovici, Larochelle, & others, 2007), and (Ranzato, Poultney, Chopra, & Cun, 2006) demonstrated that deep networks are capable of outperforming other approaches by an order of magnitude on some problems. The papers covered these three principles ("Introduction to Deep Learning Algorithms — Notes de cours IFT6266 Hiver 2010," n.d.):

1. Unsupervised learning of representations is used to train each layer.
2. Training is done on one layer at a time.
3. Supervised training is used in order to fine-tune all the layers.

In this thesis, deep convolutional neural networks are used for handling the LULC classification problems. Recently proposed architectures by Caffe such as AlexNet (Krizhevsky, Sutskever, & Hinton, 2012) and CaffeNet (BVLC | Caffe) have shown promising results with simple deep structures. Caffe's GoogleNet (Szegedy et al., 2014), on

the other hand, is a very complex deep network and performs better than the other two networks (AlexNet and CaffeNet) by a small margin. In this thesis, the experiments and conclusions of our approach are shown after evaluating the other existing prior art.

The proposed approach compares the training, validation and testing of models that are trained using transfer learning methods. The resulting classifiers are also tested on unrelated data to test the potential generalizability of labeling the data.
To compare with the existing state-of-the-art methods, the proposed approach has used UC Merced dataset that consists of 21 different classes. Each class has 100 images, and each image measures 256x256 pixels. The images were originally extracted from large images from the Unites States Geological Survey (USGS) National Map Urban Area Imagery ("The National Map: Orthoimagery," n.d.) collection for various urban areas around the country. The pixel resolution of this public domain imagery is 1 foot ("UC Merced Land Use Dataset," 2016). Please refer to Figure 6 for different image class categories. Figure 6 shows all the samples of the UC Merced dataset.

The thesis makes two main contributions:

1. Transfer learning is used to solve LULC problem using ConvNets and shown to be competitive with the state-of-the-art. However, the generalizability of a ConvNet trained using refinement learning was shown to have issues when applied to other sources of imagery of the same type used in training.
2. Dataset augmentation by rotating training images is shown to improve the generalizability of classifier built using ConvNets.

CHAPTER 2

BACKGROUND LITERATURE

Ojala et al., (Ojala, Pietikainen, & Maenpaa, 2002) have proposed a multi-resolution approach to gray-scale and rotation invariant texture classification based on local binary patterns and distributions. Lowe (Lowe, 2004) has proposed a method for extracting distinctive invariant features from images that can be used to perform reliable matching between different views of an object or scene. The features are invariant to image scale and rotation and are shown to provide robust matching across a substantial range of affine distortion, an addition of noise and change in illumination. These methods do not use convolutional neural networks or support vector machines (SVM) for their work. Dalal et al., (Dalal & Triggs, 2005) propose an approach that uses linear SVMs. They show after reviewing edge and gradient based descriptors that grids of histograms of oriented gradient descriptors significantly outperform other feature sets for human detection existing at the time.

A revolutionary work was published by Hinton et al., (Hinton et al., 2006), a fast learning algorithm for deep belief nets, that proposes a fast, greedy algorithm that can learn deep, directed belief networks one layer at a time. Bengio et al., ("Greedy Layer-Wise Training of Deep Networks - LISA - Publications - Aigaion 2.0," n.d.) (Bengio et al., 2007), which in the context of Hinton et al., continue the research that studies the algorithm empirically and explore variants to better understand its success and extend it to cases where the inputs are continuous or where the structure of the input distribution is not revealing enough about the variable to be predicted in a supervised task. Ranzato et al., (Ranzato et al., 2006) propose a novel unsupervised method for learning sparse, over complete features. These above said published works in 2006-07 have changed the course of Artificial Neural

Networks history, and they stand as an inspiration for the later developments in the Deep Learning Methods ("Convolutional Neural Networks (LeNet) — DeepLearning 0.1 documentation," n.d.).

Yang et al., (Yang & Newsam, 2010), Bag-of-visual-words(BOVW) and spatial extensions for land use classification is yet another exciting technological enhancement that uses the frequencies but not the locations of quantized image features to discriminate between classes analogous to how words are used for text document classification without regard to their order of occurrence. Their methods are evaluated using UC Merced dataset. They show an accuracy of 76.81.

Pesaresi et al., (Pesaresi & Gerhardinger, 2011) proposed an automatic recognition of human settlements in the arid regions with scattered vegetation. Their methods are based on subtraction of the vegetated areas from the built-up areas detected using the analysis of image measures extracted using anisotropic rotation-invariant gray-level co-occurrence matrix and on the introduction of a morphological filtering step. They shown an accuracy of 88.69% with morphological filtering and 70.37% with the vegetation subtraction method. Their methods do not involve neural networks, and compared to them, our approach demonstrates higher accuracy levels.

Rizvi et al., (Rizvi & Mohan, 2011) have proposed object-based image analysis of high-resolution remote sensing images using a kernel called cloud basis function, and they have investigated the probabilistic relaxation labeling process and have shown a 4% higher classification accuracy (91.47%) compared to the conventional Artificial Neural Networks existing at the time.

Yang et al., (Yang & Newsam, 2011) have proposed a spatial pyramid co-occurrence representation that characterizes both photometric and geometric aspects of an image. They show an accuracy of 77.38%.

Krizhevsky et al., (Krizhevsky, Sutskever, & Hinton, 2012) have proposed an ImageNet classification with deep convolutional neural networks. They achieved a winning top-5 test error rate of 15.3% compared to 26.2% obtained by the second best entry, with almost twice as much error rate reduction to the next best entry this paper was crucial to the research and development of the deep learning algorithms and frameworks.

Chen et al., (S. Chen & Tian, 2015) have proposed a pyramid-of-spatial-relations model to capture absolute and relative spatial relationships of local features. They employ a novel concept of spatial relation to describe a relative spatial relationship of a group of local features. They have shown that their model is robust to translation and rotation variations and demonstrates excellent performance for the application of remotely sensed land use classification. They achieve an accuracy of 89.10% which is higher than at the time state-of-the-art classification accuracy. Chen et al., (S. Chen & Tian, 2015) also proposed a spectral-spatial classification of hyperspectral data based on deep belief network. They verify the eligibility of restricted Boltzmann machine and deep belief networks and offer a novel deep architecture which combines the spectral-spatial feature extraction and the classification together to get high classification accuracy. They also propose image segmentation, which is extraction and categorization of individual features from the images, and they achieve an overall accuracy of 95.45% on image segmentation. The proposed approach in this thesis concentrates on an overall image classification rather than image segmentation.

Ren et al., (Ren, Jiang, & Yuan, 2015) proposed a new approach to tackling high-dimensional local binary patterns. Their objective was to select an optimal subset of binarized-pixel-difference features to compose the local binary pattern structure. They take the advantage of the fact that the local features are closely related, and they propose an incremental Maximal Conditional Mutual Information (MCMI) scheme which learns local binary patterns. Their approach shows an accuracy of 88.20%.

Hu et al., (Hu et al., 2015) present an improved unsupervised feature learning algorithm based on spectral clustering that adaptively learns good local feature representations and also discovers intrinsic structures of local image patches. The approach first maps the original image patches into a small dimensional and inherent features space by linear manifold analysis techniques and then learns a dictionary using K-Means clustering on the patch manifold for feature encoding. They experimented with the UC Merced dataset and have shown to achieve 90.26% accuracy rate.

Shao et al., (Shao, Yang, Xia, & Liu, 2013) propose a classification model based on a hierarchical fusion of multiple features. They employ four discriminative image features and an SVM with histogram intersection kernel in different classification stages, they conduct an extensive evaluation of various configurations and show an accuracy rate of 92.38%. Negrel et al., (Negrel, Picard, & Gosselin, 2014) present an investigation that uses visual features based on second-order statistics, as well as new processing techniques to improve the quality of features. They experiment on UC Merced dataset and show an accuracy of 94.30%.

In their survey paper on transfer learning, Pan et al., (Pan & Yang, 2010) point out that conventional classification approaches have assumed that training data and future inputs to a classifier must be statistically similar. However, significant problems can be solved by knowledge transfer from one task (where a large amount of data is available) into another domain with much fewer examples. Deep networks can make transfer learning especially attractive (Bengio, 2012) because feature modeling and higher order knowledge tend to live at different layers. Recently, Castelluccio et al., (Castelluccio, Poggi, Sansone, & Verdoliva, 2015) have shown that excellent classification results can be obtained by using a pre-trained classifier tuned for a larger dataset (ImageNET) and refining it by replacing the last layers. They show 97.1% accuracy on UC Merced dataset by improving a pre-trained classifier GoogleNet (Szegedy et al., 2014), but the performance of this model was not tested on unrelated data. Catelluccio et al., show the results with five-fold validation (80 – train and 20 - validate) without separating out any data for testing the resultant classifier (the resultant classifier is the final model after it has been trained). The proposed approach in this thesis extends it by using a testing dataset along with the training and validation dataset (60 - training, 20 - validate and 20 - testing) and also verifies the performance of the classifier on wholly unrelated data. This helps us in building a classifier that can categorize large amounts of unseen new data.

Deep learning techniques have been very recently applied to the LULC problem, Lv et al., (Lv et al., 2015) used deep belief networks for LULC on radar imagery. Unlike the proposed solution in this thesis, their approach is not applied to RGB imagery and it not convolutional. Panetti et al., (O. A. B. Penatti, Nogueira, & Santos, 2015) experimented with convolutional networks for aerial and remote sensing images, finding that a network trained for visual RGB data outperformed all other methods on the aerial images. However, low level descriptors, such as BIC (Border Interior Pixel Classification), proposed by Stehling et al., (Stehling, Nascimento, & Falcão, 2002) outperformed the pre-trained convolutional neural networks on remote sensing images. Castillucio et al. (2015) improve on these results by retraining GoogleNet's classifier by fine-tuning on the same data to outperform all other approaches. The approach in this thesis has achieved similar results as Castillucio et al.

(2015). Unaware of each other, our approach in this thesis and the work of Castillucio et al. were experimented at the same time.

Romero et al., (2016) (Romero, Gatta, & Camps-Valls, 2016) very recently proposed an unsupervised deep feature extraction for remote sensing image classification. They suggest the use of greedy layer-wise unsupervised pre-training coupled with an algorithm for unsupervised learning of sparse features. They show an average accuracy of 74.34%, which is a decent accuracy for an unsupervised classifier. Pre-training approaches for unsupervised deep networks is currently active and an important research area.

CHAPTER 3

METHODOLOGY

Convolutional Neural Networks (CNN) are made up of perceptron's that are biologically analogous to the neurons of a human brain. Each perceptron has learnable weights and biases. A dot product of the inputs and their corresponding weights is taken, and a non-linear function is applied to produce the corresponding output. The output of one perceptron is fed into the perceptron of another layer, and this process continues and during this process, the weights and biases of each layer are learned incrementally. The input and the output are the outer layers of a neural network. Each layer applies a function to the output of the previous layer. Hidden layers are the in between layers that do the computation. The hidden layers' job is to apply convolution and pooling operations alternatively and adjust the weights and biases of the network according to the input.

In a regular neural network popularly called as Multi-Layered Perceptron (MLP), each perceptron is fully connected to all the perceptrons of the next layer as shown in Figure 2(*Left*). MLPs are manageable for small-scale images, for example, a 28x28 RGB image would require 28x28x3 weights per perceptron to be computed. The problem with MLP arises with a large scale image as a large number of weights have to be learned, for example, a 256x256 RGB image would need 256x256x3 weights to be computed, moreover one would want to have several such neurons, this will result in a humongous allocation of computational time and resources. Adding to that, learning so many parameters would lead to a problem called Over-fitting. Unlike regular MLPs, CNN's are biologically inspired variants of MLPs ("Convolutional Neural Networks (LeNet) — DeepLearning 0.1 documentation," n.d.). Convolutional Neural Networks have sparse connectivity, for example in Figure 2(*Right*) each perceptron takes input from three perceptrons from the previous layer. These group perceptrons are also called activations. CNN's take advantage of the input images by localizing the reception of features (features in image are non-dynamic, and are spatially close to each other). This process exploits the spatially-local correlated contiguous fields called receptive fields (Hubel & Wiesel, 1968) by enforcing a local connectivity pattern between neurons of adjacent layers ("Convolutional Neural Networks (LeNet) — DeepLearning 0.1 documentation," n.d.). Moreover, all neurons of a layer are identical to one another, except for their receptive fields, sharing the same weights. This reduces the number of weights to be learned. From the Hubel et al., (Hubel & Wiesel, 1968) work on cat's visual cortex we know that convolutional neural network closely resembles the biological visual cortex, which is organized in layers composed of similar cells, with different receptive fields over the layers.

Convolutional neural networks primarily use the following different types of layers:

1. Input Layer: Contains the input image or the raw pixel values. It is the entry layer to all the other layers.
2. Convolutional Layer: This layer computes the activations of perceptrons that are connected to the receptive fields of the previous layer. As discussed above, each perceptron is connected to a spatially local region of the input volume. The convolutional parameters of a layer include:
    a. Number of outputs, is the input for the next layer.
    b. Kernel size, controls the spatially local region of the input volume.
    c. Stride, the pixel skips of the sliding window.
    d. Padding, helps in sizing the layer.

3. Pooling Layer: This layer is mainly used to resize and accumulate the spatial representations. For example, using a max() operation is called max pooling. It is quite common to sandwich a pooling layer between convolutional layers periodically.
4. Normalization Layer: Normalizes over local input regions which helps in generalization.
5. Fully-connected Layer: These are typically the last couple of layers of the network. Perceptrons in a fully connected layer are fully connected to all activations of the previous layer. The difference between a fully connected layer and a convolution layer is that the perceptrons in the convolution layer are connected only to a local region in the input, whereas all the perceptrons in the fully connected layer are connected to all the perceptrons of the input (input to the fully connected layer).

Figure 4. shows each of the 96 filters (edge detections) that are learned in the first layer of a CaffeNet. Dumitru et. al, (Dumitru Erhan, n.d. "Understanding Representations Learned in Deep Architectures - LISA - Publications - Aigaion 2.0," n.d.) propose a paper to find good qualitative interpretations of high-level features represented by CNN models. These representations are patterns that can be displayed and are meaningful to the human eye.

Caffe created by Yangqing Jia is one of the popular frameworks for the deep learning methods. It is developed by Berkeley Vision and Learning Centre (BVLC) and community contributors. The main reason for us to use Caffe framework is because it enables us to extensively utilize the GPU for hardware acceleration, and it also provides a framework to build, edit and run custom networks for refining. Caffe can also be run on CPU only. The highly customizable features of Caffe through configuration, prototxt and solver files makes it easy to use and concentrate more on the research techniques instead of worrying about implementing the deep architectures. The code is released under BSD 2-Clause license, and it can be forked and customized to satisfy ones' research requirements. It also provides C++, Python and Matlab APIs for easy accessibility. Figure 3. shows a typical architecture of the CaffeNet. It consists of five convolutional layers, each followed by a pooling layer, and three fully-connected layers.

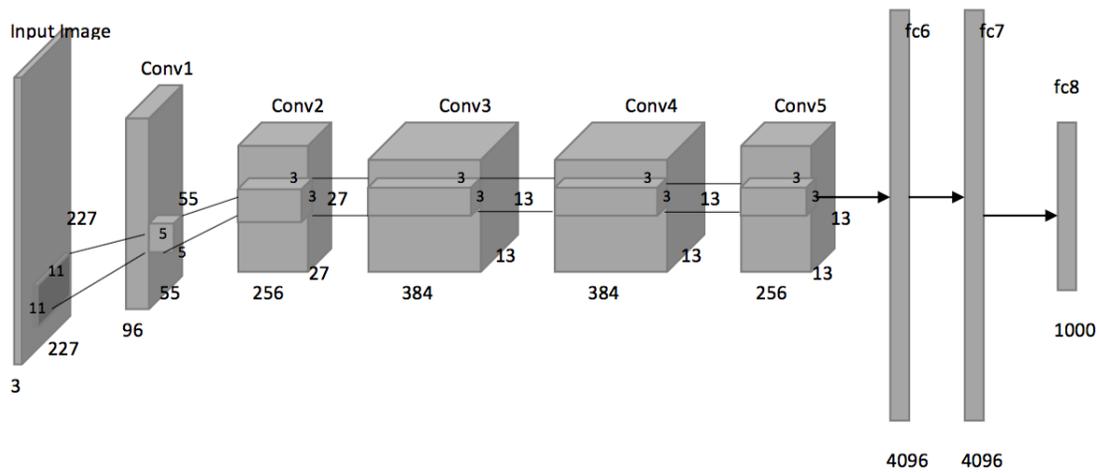

*Figure 3*. The original CaffeNet architecture used in this work. fc8 layer from this architecture is modified (number of outputs is 21 instead of 1000) to classify 21 classes

Alex Krizhecsky et al., (2012) (Krizhevsky, Sutskever, & Hinton, 2012) have trained a large, deep convolutional neural network to classify 1.2M high-resolution images in the ImageNet LSVRC-2010 contest into 1000 different classes and this model file is available at Caffe's model zoo as AlexNet. CaffeNet is a minor variation from the AlexNet with few differences in network topology, data preparation, and augmentation and averaged classification. In the proposed work of this thesis, we have changed the last layer to limit the classifications to 21 classes. So instead of 1000, we have 21 categories in the output layer and these 21 classes are the classes from UC Merced dataset. CaffeNet's Conv1 as shown in Figure 3. is the only layer that is exposed to the raw image. The 96 filter weights of Conv1 are visualized in Figure 4. Visualization of the layers helps in verifying the well trained networks, which usually display smooth filters without any noisy patterns (Alizadeh & Fazel, n.d.).

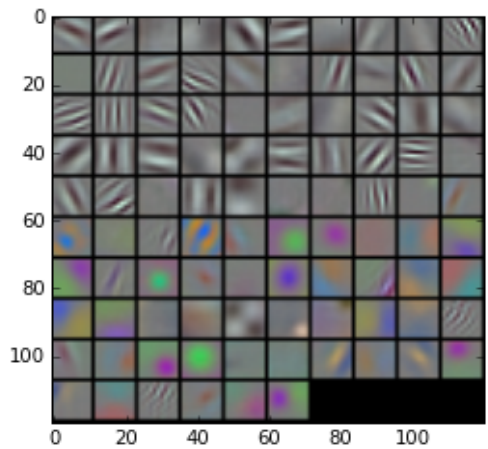

*Figure 4*. Typical looking filters of Conv1 layer

GoogleNet, on the other hand, is as exactly as described by Szegedy et al., (2014) (Szegedy et al., 2014) in Going Deeper with Convolutions paper. GoogleNet is rather a deep and complex network compared to the CaffeNet or the AlexNet. GoogleNet won the ImageNet Large-Scale Visual Recognition Challenge 2014 - ILSVRC14 challenge ("ImageNet Large Scale Visual Recognition Competition (ILSVRC)," n.d.). It is a 22-layer deep network, and according to Szegedy et al., the the primary hallmark of its architecture is the improved utilization of the computing resources inside the network, a network in network module derived from Lin et al., (2013) (Lin, Chen, & Yan, 2013). This module is termed as Inception module. A snapshot of the inception module is shown in Figure 5.

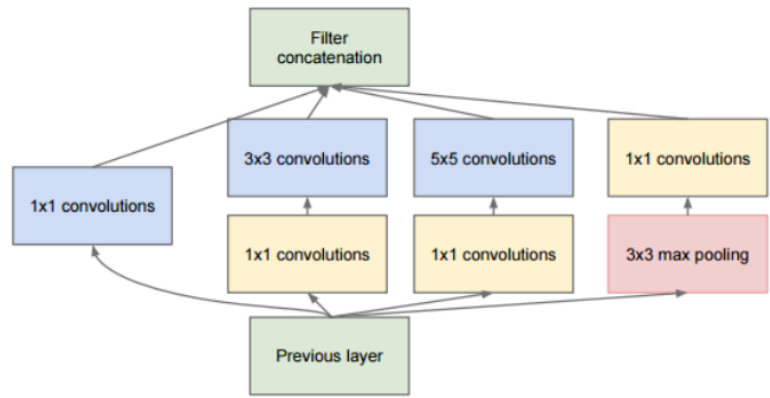

*Figure 5.* Inception module with dimension reductions

These inception modules are stacked on one another, thus enabling higher layers to capture features of higher abstraction. For complete description of the 22-layer architecture, please refer to Szegedy et. al. (2014) (Szegedy et al., 2014).

The architectures discussed above have been developed to process RGB images to find patterns and extract prominent features and correlate these features with the learned weights to update the weights of the system. Learning from images is a standard and an easy task for human beings to perform. But neural nets need large datasets to train and perform better, and training on large datasets is a time taking process (heavily depends on the hardware acceleration). A typical network may take days/ weeks to train from scratch depending on the hardware capabilities used. Krizhevsky et. al, (2012) (Krizhevsky et al., 2012) claim that the dataset takes 5-6 days to train AlexNet on two NVIDIA GTX 580 3GB GPU's.

In remote sensing, one has access to massive amounts of data, but most of this data is unlabeled. To categorize these remote sensing images, one needs a LULC classifier. But, to train this classifier, one runs into a very limited labeled training data. It is quite a challenge to train a classifier from scratch on these small datasets and obtain good classification accuracy levels. As discussed above, a common problem that we observe by training on small datasets is over-fitting. The classifier works with the best accuracy levels on the training data, but it does not generalize well to test data. The proposed approach in this thesis tries to address this problem, and experiments are conducted on unseen/ unrelated San-Diego remote sensing data to verify the same. The proposed method also evaluates Brazilian Coffee Scenes dataset which is a peculiar dataset much different from the ImageNet dataset.

Our motivation is Razavian et al., (2014) (Razavian, Azizpour, Sullivan, & Carlsson, 2014), Jia et al., (2104) (Jia et al., 2014), and Yosinski et al., (2014) (Yosinski, Clune, Bengio, & Lipson, 2014), as they have explored this possibility, published state-of-the-art results and suggested that the features obtained from deep learning with convolutional nets are to be considered as the primary candidate in most visual recognition tasks.

The proposed approach in this thesis takes the advantage of the availability of existing deep network models and the ability to quickly train them by transfer learning and fine-tuning to generate state-of-the-art classifier that performs and generalizes better than most of the prior-art classifiers.
Transfer learning can be categorized into two variations, our approach presents the experiments and results based on these variations:

1. SVM (Support Vector Machines): Feature sets from the top most layers of Caffe, Alex or Google Net are taken and are trained on an external classifier (Support Vector Machine – SVM, in this case). The features that are borrowed are generally from fully connected layer features and activations of each perceptron of this layer is dependent on all of the perceptrons of the previous layer. This helps in non-localization of the classification.
2. Fine-tuning: Fine tune by replacing the last layer and updating the weights of the pre-trained network by changing the base learning rate and the learning rate of the other layers to continue back propagation. The learning rate of other layers can be set to zero if the pre-trained network weights are to be untouched. The learning rates are adjusted in such a way that the last layers learn faster compared to that of the other layers.

In this thesis, three existing frameworks are considered, AlexNet, CaffeNet and GoogleNet and performances of these are compared using Caffe framework. NVIDIA's Deep Learning GPU Training System (DIGITS) is yet another framework that enables us to harness the power of DNN's running in parallel on multi-GPU systems. In this thesis, TESLA K40c 12 GB and QUADRO K4000 3GB GPUs have been used for hardware acceleration.

CHAPTER 4

DATA ANALYSIS AND RESULTS

Experiments were conducted on three datasets: 1. UC Merced land use scenes, 2. Brazilian Coffee scenes and 3. San Diego data. UC Merced dataset, as discussed in previous chapter, consists of 21 land-use classes selected from aerial orthoimagery. Each set contains 100 images measuring 256x256 pixels for each of the 21 categories as shown in Figure 6. These classes include a variety of spatial patterns, some homogeneous on texture, some homogeneous on color and others not homogenous at all (Cusano, Napoletano, & Schettini, 2014). These diverse variations of the dataset make it a good experimental dataset. UC Merced dataset represents similar spatial characteristics to that of the ImageNet (Castelluccio et al., 2015). The Brazilian Coffee scenes (O. A. B. Penatti et al., 2015) dataset, on the other hand, include satellite images with an infra-red band, and these are less similar to that of ImageNet dataset. The dataset is categorized into Coffee and Non-Coffee scenes. Each image of Brazilian dataset measures 64x64 pixels. The San Diego dataset is specially downloaded from the United States Geological Society ("EarthExplorer," n.d.), earth explorer website, to verify the generalizability of the classifier. An aerial patch was selected (from earth explorer website) in a way that it consists of as many classes as possible from the UC Merced dataset.

1. UC Merced

    The images from this dataset share many low-level features with that of ImageNet, and this is the prime reason for this dataset to perform consistently and exceptionally well while fine-tuning the pre-trained networks. Table 1. shows the classification accuracy levels of the proposed solutions. Fine-tuning GoogleNet gives an accuracy

of 96%. AlexNet and CaffeNet also show accuracy levels of above 95%. Table 2, shown below, compares the prominent approaches that were built to achieve the state-of-the-art precision. Castelluccio et al., (2015), propose the highest accuracy level with 97.1, a positive difference of 1.1 as compared to the highest accuracy (96) reported by us.

| CNN | Approach | Accuracy |
|---|---|---|
| AlexNet | Fine-tuning | 95.79 |
|  | svm | 94.76 |
| CaffeNet | Fine-tuning | 95.02 |
|  | svm | 92.85 |
| GoogleNet | Fine-tuning | **96.0** |
|  | svm | 94.28 |

*Table 1.* Classification accuracies levels of the proposed solutions

All the validations that are shown in Table 1 are five-fold validations (80 - Train, 20 - Validate), this is done to compare the accuracy levels with the other state-of-the-art classifiers. Note that all the fine-tuning models show an accuracy level of 95% and above. AlexNet, CaffeNet, and GoogleNet are fine-tuned with 25,000 iterations. To train a Support Vector Machine (SVM), the output of the fully connected layer 7 is used as the input for the SVM in the case of AlexNet and CaffeNet, and the output of the penultimate layer of GoogleNet is used as the input for the SVM in the case of GoogleNet. Scikit-learn's SVM is used to achieve these accuracy levels.

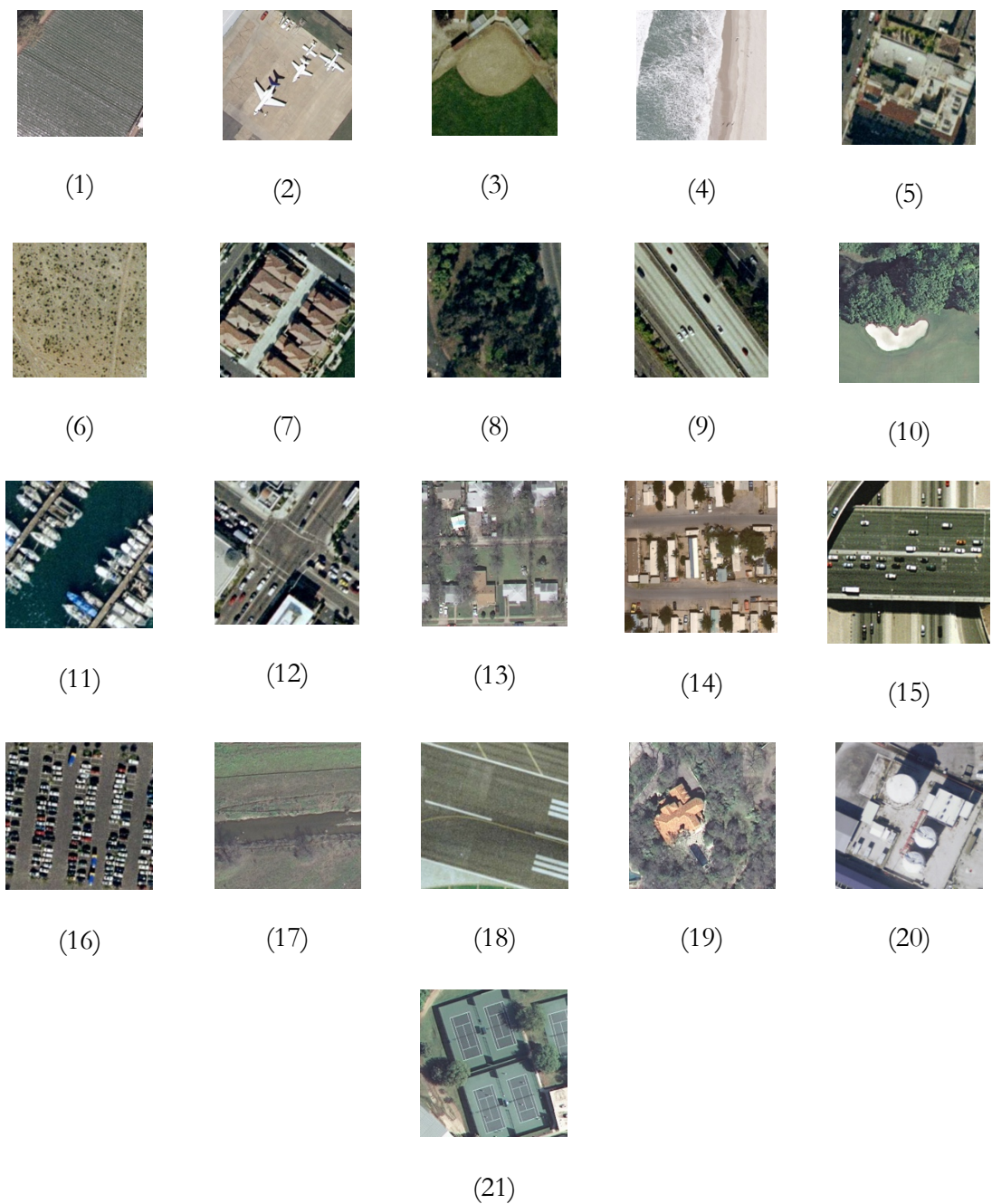

*Figure 6.* Example images associated with 21 land use categories in the UC Merced dataset. (1) Agricultural. (2) Airplane. (3) Baseballdiamond. (4) Beach. (5) Buildings. (6) Chaparral. (7) Denseresidential. (8) Forest. (9) Freeway. (10) Golfcourse. (11) Harbor. (12) Intersection. (13) Mediumresidential. (14) Mobilehomepark. (15) Overpass. (16) Parkinglot. (17) River. (18) Runway. (19) Sparseresidential. (20) Storagetanks. (21) Tenniscourt.

| Year | Accuracy | Authors | Approach |
|------|----------|---------|----------|
| 2010 | 76.81 | Yang et. al. | BOVW |
| 2011 | 76.05 | Yi Yang et. al. | SPCK |
| 2013 | 92.38 | Shao et. al. | HFMF |
| 2014 | 94.30 | Negrel et. al. | VFS |
| 2015 | 88.20 | Ren et. at. | MCMI |
| 2015 | 89.10 | Chen et. al. | PSR |
| 2015 | 90.26 | Fan Hu et. al. | Unsupervised |
| 2015 | 97.10 | Castelluccio et. al. | Supervised |
| 2016 | 74.34 | Romero, et. al. | Unsupervised |
| 2016 | **96.0** | *proposed* | Supervised |

*Table 2.* Classification accuracy comparison table on UC Merced dataset, please refer to the prior-art chapter for information on the approach used by these methods.

2. Brazilian Coffee Scenes

    As discussed above these images are quite different from that of the training data (ImageNet) used to produce CaffeNet, AlexNet and GoogleNet models. The pre-trained models usually take 256x256 size images as input, but the Brazilian Coffee Scenes dataset consists of 64x64 size images. Given the said variations, our approach is to test transfer learning methods on this dataset as this is a remote sensing dataset used for similar land classification purposes. Figure 7 shows some samples of

Brazilian Coffee Scenes. The results shown below are obtained after five-fold cross validation.

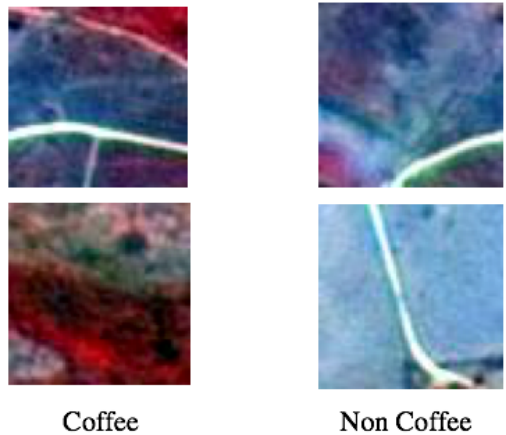

*Figure 7*. Brazilian Coffee Scenes dataset showing coffee and non coffee samples

In fine-tuning, each layer convolves into a set of filters as shown in Figure 1. Conv1 has 96 output features, Conv2-layer 2 has 256, Conv3-layer3 has 344 and so on. As the size of the original input (64x64) is comparatively smaller, the filters get nullified gradually as the features travel deep into the layers. Up-sampling helps in scalability and usage of these images for the experiment. Each image of the dataset has been up-sampled from 64x64 to 256x256 pixels using nearest neighbor resampling filter. Fine-tuning by GoogleNet after up-sampling gives the highest average accuracy of 94.1 when compared to the recent state-of-the-art results by Castellucio et. al., (2015) - GoogleNet/from-scratch. The transfer learning on SVM approach show decent results, as shown in Table 3. CaffeNet feature vectors of FC7 layer trained on SVM shows the accuracy of 85.21%. The standings of these results are presented in Table 4 - Table referenced from Castelluccio et al., (2015) and Penatti et al., (2015).

| CNN | Approach | Accuracy |
|---|---|---|
| AlexNet | Fine-tuning | 89.84 |
| | svm | 83.82 |
| CaffeNet | Fine-tuning | 91.26 |
| | svm | 85.21 |
| GoogleNet | Fine-tuning | **94.1** |
| | svm | 79.82 |

*Table 3*. Classification accuracies with transfer learning approach on Brazilian Coffee Scenes dataset.

| Method | Authors | Accuracy |
|---|---|---|
| BIC | Penatti et. al. | 87 |
| Overfeat + Caffe | Penatti et. al. | 79.01 |
| GoogleNet/from scratch | Castelluccio et. al. | 91.8 |
| Overfeat | Penatti et. al. | 83.04 |
| Transfer learning/ up-sampling/ fine-tuning | *Proposed* | **94.1** |

*Table 4*. Classification accuracy comparison table on Brazilian Coffee Scenes dataset

3. San Diego dataset

   As discussed above, the San Diego dataset is obtained by choosing a random patch from the USGS Earth Explorer website. These images were manually segregated for testing the extent of generalizability of the classifier. The fine-tuned classifier models that report high accuracy levels in Table 1. are taken and tested on this San Diego dataset. The pre-trained models provided by Caffe are used to produce SVM results. The models are tested on 661 un-seen, and unrelated San Diego images and the average accuracy results of fine-tuning and SVM approaches are reported in Table 5.

| CNN | Approach | Accuracy |
|---|---|---|
| AlexNet | Fine-tuning | 79.49 |
|  | svm | 75.94 |
| CaffeNet | Fine-tuning | 79.08 |
|  | svm | 78.88 |
| GoogleNet | Fine-tuning | **85.59** |
|  | svm | 80.39 |

*Table 5.* Classification accuracies on the San Diego dataset without dataset augmentation.

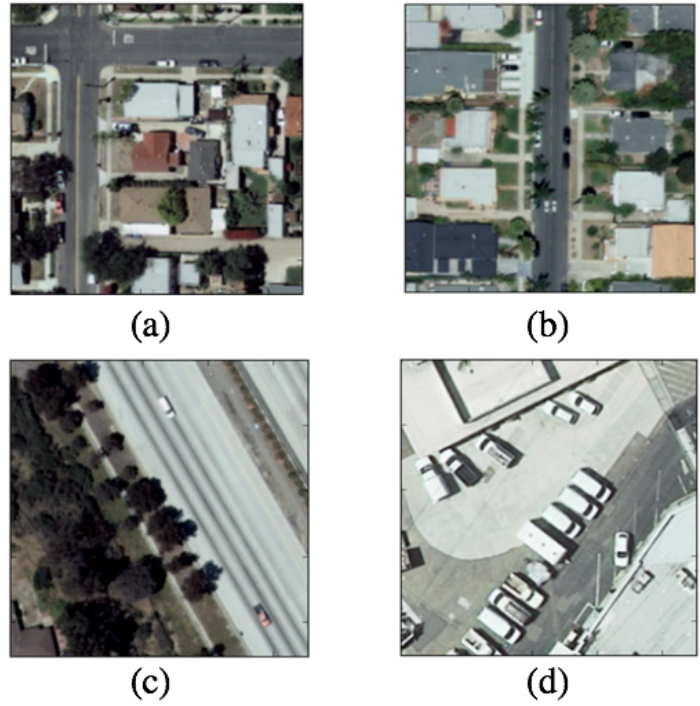

*Figure 8.* Some of the error samples found on San Diego dataset using fine-tuning approach. (a) Residential → Mobilehomepark (b) Residential → Mobilehomepark (c) Freeway → Overpass (d) Parkinglot → Storagetanks.

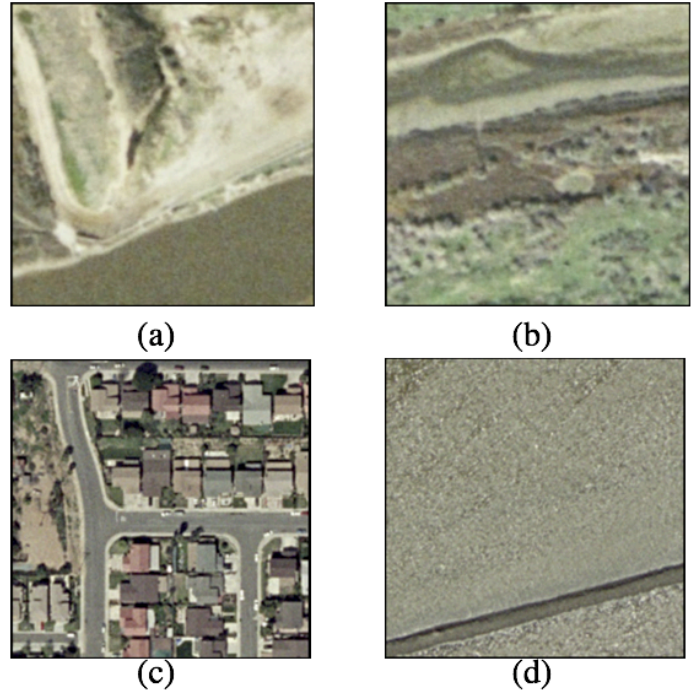

Figure 9. Some of the error samples found on San Diego dataset using SVM approach. (a) River → Golfcourse (b) River → Forest (c) Residential → Intersection (d) River → Agricultural.

4. UC Merced data augmentation

The proposed approach also experiments on augmenting the dataset by rotating and cropping each of the images of the UC Merced dataset. The dataset is divided as 60% - training, 20% - validation and 20% – testing (testing the classifier after it is thoroughly trained) per class. Each of the UC Merced dataset images is rotated by +- 5, 10, 30 and 40 degrees and added to the augmented dataset. A total of 11340 training images were generated, and these images were used as the training data for CaffeNet fine-tuning approach. Confusion matrix of the results is as shown in Figure 7. The accuracy of the resultant classifier is 85.71%. Some of the categories are closely related (from Figure 6 and Figure 10, we can say that class G - dense residential and class M - medium residential categories are closely related). These classes have subtle differences that are difficult to be differentiated even with the human eye. If these areas (class G and M, Figure 10) are combined, then the accuracy of the classifier on unseen UC Merced dataset boosts up to 89.52%.

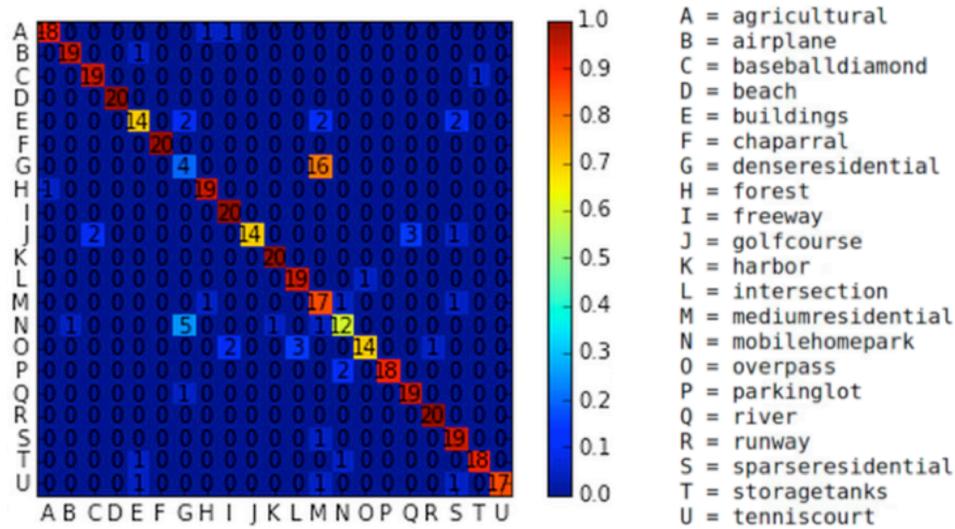

*Figure 10*. The confusion matrix of the classified unseen UC Merced testing dataset

To further investigate the data augmentation approach, classifiers are built using 80% of the augmented images for training and the remaining 20% for validation. The training dataset (UC Merced, after data augmentation) consists of a total of 15120 images. The resultant classifiers are tested on San Diego data to verify the generalizability feature. The highest accuracy achieved by this process is 88.07% by GoogleNet as shown in the Table 6. We can observe from Table 6 that classifiers obtained with data augmentation perform consistently even on the unseen/

unrelated San Diego data. However, AlexNet shows a slight indifferent behavior when compared to CaffeNet or GoogleNet, as accuracy on San Diego data with augmentation does improve over the accuracy obtained with out data augmentation. To verify this behavior, the classifier (obtained with augmentation) is tested on augmented San Diego dataset (augmented as described above), the average accuracy obtained by this process is 79.83%, which is slightly higher than 79.49% (accuracy w/o data augmentation). This experiment shows that the dataset augmentation by image rotation improves the generalizability of classifier built using CNNs.

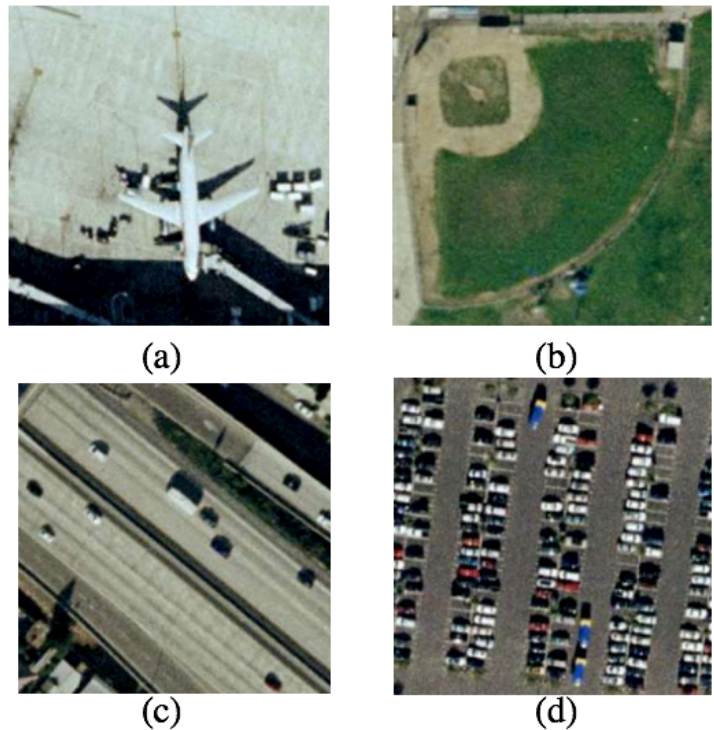

*Figure 11*. Error samples by the classifier (CaffeNet) built using the data augmentation approach (a) Airplane → Buildings (b) Baseballdiamond → Sparseresidential (c) Freeway → Overpass (d) Parkinglot → Harbor

| CNN | Dataset | With Augmentation | Without Augmentation |
|---|---|---|---|
| AlexNet | UC Merced | 80.71 | 95.79 |
| | San Diego | 78.4 | 79.49 |
| CaffeNet | UC Merced | 82.54 | 95.02 |
| | San Diego | 81.63 | 79.08 |
| GoogleNet | UC Merced | 88.58 | 96.0 |
| | San Diego | **88.07** | 85.59 |

Table 6. Classification accuracies on the UC Merced and San Diego datasets with/ without dataset augmentation.

Filter Visualizations:

Filter visualization is a way of knowing whether a model learned the correct parameters or not by visual confirmation. Recent developments in filter visualizations ("Research Blog: Inceptionism: Going Deeper into Neural Networks," n.d.) has been a huge help for us in determining the particular categories on which the GoogleNet model has to be further trained or fine-tuned. Figure 12, shows two different classes of the UC Merced dataset (Harbor and Medium residential), (a) shows the original image, (b) shows the image as seen by the pre-trained GoolgeNet model, and (c) shows the image as seen by the fine-tuned GoogleNet model (as proposed in this thesis). The Harbor category is well trained, and its accuracy is close to 100%, whereas the Medium-residential class is often confused with other classes, such as Dense-residential, Sparse-residential and Mobile-home-park. If we compare the fine-tuned images ((c)) of both the categories we can see extra artifacts in the Medium-residential visualization (roof tops as human faces, over-fitting) when compared to that of Harbor. We can deduce by looking at these visualizations that the fine-tuned GoogleNet model is well trained on Harbor (as it shows less/ nil artifacts), but the model needs further fine-tuning/ training on the Medium-residential category.

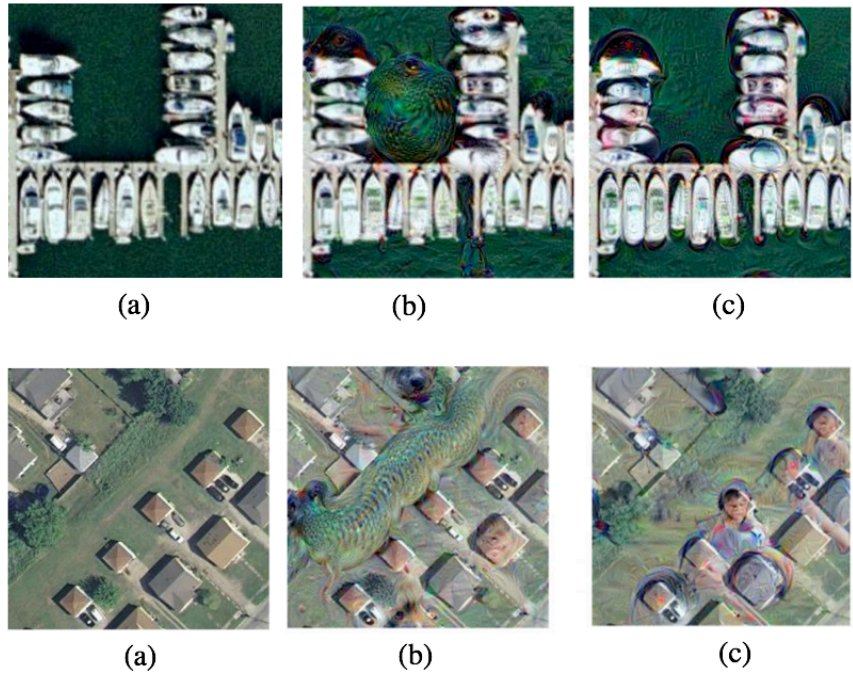

*Figure 12*. Filter visualizations of Harbor(*top*) and Medium-residential(*bottom*)

CHAPTER 5

DISCUSSION

Deep learning has had a transformative effect on computer vision. The proposed approach in this thesis has shown that it can be applied to remote sensing applications for automatic LULC classification from VHR images. The approach demonstrated that networks trained on an unrelated image recognition task can actually be used to solve the LULC classification problem. One would anticipate that a large amount of VHR spatial imagery that already exists and that continues to be collected at higher rates will have a significant impact on a variety of remote sensing applications. The proposed approach has shown two transfer learning methods, 1. Fine-tuning and 2. Feature vector on SVM. Both the methods show accuracies that are at par with the state-of-the-art accuracies on the LULC classification problem. The approach also deduced that these methods are consistent on a similar type of the datasets as the original dataset used for training (ImageNet), this is shown by experimenting with Brazilian Coffee Scenes dataset. The San Diego data classification clearly shows us that the proposed classifiers generalize well on completely unseen data. This proposed approach is a breakthrough as significant unlabeled remote sensing datasets can now be classified and categorized.

Adapting a deep pre-trained network and fine-tuning the network on a new dataset that has a limited number of labeled images to train quickly, learn and adjust the weights and biases of the network on the new dataset in effect delivers promising results. From Table 2, one can deduce that the other non-CNN methods are at a wide 6% below the proposed accuracy. The near perfect accuracy levels from Table 1 shows that this is the best bleeding edge solution to LULC problem. The next big challenge would be pixel level extraction of cars, trees and other prominent features from the remote sensing datasets. Drones have also become widespread resources for the aerial imagery, we intend to work on these images too in the future for LULC and pixel level image segmentation and feature extraction.

APPENDIX A

TREE EXTRACTION

Now that the land use and land cover classification is evaluated successfully, future plans are to extend the research to the extraction of trees from aerial/ remote sensing images using deep learning methods. Trees are significant and important features in outdoor scenes, and they can have a tremendous impact on simulation, training, and scientific modeling efforts.  In particular, the relationships between tree placement and cultural features such as buildings in urban environments is important because they interact with people and architecture. Trees are important for urban planning applications because the shade cast by trees may impact the walkability of an environment, and also affect the amount of water and energy needed in desert cities.  Tree placement is also important for ground based games or simulation applications for realism and immersion, and also because of the features occluded by trees and the cover they provide. In military or unmanned aerial system (UAS) sensor operator simulations, trees can provide cover and hide targets.

Accurate data on the placement of trees can be expensive to produce. Often tree cover can be estimated by extracting foliage from high-resolution Light Detection and Ranging (LiDAR) or stereophotogrammetry data, but trees change quickly compared to other elements like buildings, and it is currently difficult to maintain enough complete and current source material. Recent trends in sensors and commercial applications for satellite and aerial imagery have made orthophoto mosaics at resolutions of around one meter per pixel or less more accessible than ever before.  Algorithmic solutions can place statistically likely trees in urban environments, but randomly placed trees may not capture the complicated relationships between cultural features like buildings and trees. There is a need for solutions that can algorithmically place trees guided by existing aerial photographs so that a procedurally generated 3D scene is visually consistent with orthophoto and other mapping data available for a view.

The primary challenge for tree extraction is that the range of colors in trees overlaps the range of colors for other vegetation such as grass. Texture attributes can improve the ability to separate, but the texture of foliage can be obscured by image under-sampling or compression artifacts. Shadows are often a good indication of the presence of a tall feature such as a tree, and shadows also appear to give an indication of the height and shape of a tree. However, in near-nadir viewpoints treetops, there may not be a clear separation between tree top and shadow. Trees are often surrounded by features which obscure or hide their shadow, and the foliage of trees is rough, and so branches of an individual tree cast shadows within the treetop itself. Future work would extend to the auto extraction of individual features from remote sensing images thereby segmenting, separating and representing different pixels of these features clearly to the human eye.

BIOGRAPHICAL SKETCH

Nagesh K Uba is currently in his 2nd year of study in the Software Engineering Program at the Arizona State University, Polytechnic Campus. In May 2016, he will graduate with a Master of Science in Software Engineering degree, with a focus in Deep Learning Techniques. Before ASU, Nagesh worked as a Software/Application Developer at Tata Elxsi and Oracle Corporation companies for more than five years. Nagesh received his Bachelor's degree in Computer Science and Engineering from Visvesvaraya National Institute of Technology, Nagpur – India. While at the ASU, he has served as a Teaching Assistant in the Department of CIDSE at the ASU, Polytechnic.